\documentclass{article}
\usepackage{spconf,amsmath,graphicx, url, xcolor, amsfonts}

\newcommand{\T}[1]{\mathcal{T}_\mathrm{#1}}
\title{Tempo vs. Pitch: understanding self-supervised tempo estimation} 
\name{$\begin{array}{cc} \mbox{Giovana Morais$^{1}$
, Matthew E. P. Davies$^{2}$, Marcelo Queiroz$^{1}$
, Magdalena Fuentes$^{3}$}\end{array}$ }
\address{$^1$ Computer Science Dept., University of S\~{a}o Paulo, S\~{a}o Paulo, Brazil\\\
$^2$ SiriusXM, USA\\
$^3$ MARL-IDM, New York University, New York, USA}

\begin{document}
\ninept
\maketitle
\begin{abstract}
Self-supervision methods learn representations by solving pretext tasks that do not require human-generated labels, alleviating the need for time-consuming annotations. These methods have been applied in computer vision, natural language processing, environmental sound analysis, and recently in music information retrieval, e.g. for pitch estimation. Particularly in the context of music, there are few insights about the fragility of these models regarding different distributions of data, and how they could be mitigated. In this paper, we explore these questions by dissecting a self-supervised model for pitch estimation adapted for tempo estimation via rigorous experimentation with synthetic data. Specifically, we study the relationship between the input representation and data distribution for self-supervised tempo estimation.

\end{abstract}
\section{Introduction}
\label{sec:intro}

Tempo is a fundamental dimension of music related to the speed at which a listener would tap along to mark the underlying pulse. Automatic tempo estimation has received a great deal of attention in the last decade and it is still considered a challenging task because of tempo's intrinsic ambiguity, given the various possible interpretations of the metrical structure of a given rhythm, and because of weak note onsets, syncopation, and local tempo changes~\cite{schreiber2020}.
Tempo estimation received considerable attention in the Music Information Retrieval (MIR) community due to its many potential applications \cite{schreiber2018single, foroughmand2019deep, bock2020deconstruct}, such as automatic playlist generation and recommendation, or as the input for other MIR tasks such as transcription or synchronization \cite{peeters2006template}. Similarly to how pitch is a reference for harmonic sounds, tempo serves as a reference for periodicity in the context of rhythm \cite{peeters2006template}.

While early methods were largely based on signal processing, combining techniques such as autocorrelation or Fourier analysis with peak picking \cite{peeters2006template, zapata2011}, recent approaches exploit deep neural networks (DNNs). Similarly to other MIR tasks, the introduction of DNNs meant a change of paradigm and improvement in the accuracy of tempo estimation models. New methods exploring convolutional neural networks (CNNs) \cite{schreiber2018single}, and more recently temporal convolutional networks \cite{bock2020deconstruct} became state of the art.

Although DNNs led to a significant improvement in automatically estimating tempo, it is well known that these models are very data-hungry \cite{bock2020deconstruct}. This is particularly problematic in the context of MIR, where obtaining annotated data is expensive, as producing labels often requires musical expertise. In other fields, such as computer vision, this data bottleneck inspired the use of \mbox{\textit{self-supervision}}, a set of methods where labels are derived algorithmically from the data without the need for human supervision. In self-supervision, models are trained to solve a \textit{pretext task}, which might not be relevant for the final application for which the model will be used (\textit{downstream task)}, but exploits some intrinsic information of the data. If designed well, by solving the pretext task the model learns meaningful internal representations that can be extrapolated to the problem of interest. This approach has been exploited in computer vision, environmental sound analysis, and very recently, in MIR for pitch \cite{spice2020} and beat tracking \cite{zerosamba2022}. 


A major drawback of DNNs, which has not improved with the introduction of self-supervision, is the lack of intuition behind what models are learning and how fragile they are with respect to different training data distributions. For example, it has been shown that models trained on imbalanced distributions in self-supervised settings transfer this bias to downstream tasks, making it difficult to revert this behavior at later stages \cite{wu2021exploring}. In this way, understanding the interplay between the data distribution and the model can help inform decisions about which data to use for training and how to implement data augmentation.

In this paper, we explore these questions by adapting a self-supervised model for pitch estimation \cite{spice2020} (SPICE) to tempo estimation, inspired by the similarities between the two tasks. Particularly, using synthetic data we look at the interplay between the input representation and the data distribution and its effect on the performance of this model. We then discuss the implications in the robustness of this method and its adaptability to real data.

\section{Related Work}
\label{sec:relatedwork}

Desblancs et al.~\cite{zerosamba2022} investigated self-supervised beat tracking, for which they trained two different CNNs, one to process the percussive content and another to process the harmonic content of a signal, where the networks solved the pretext task of synchrony prediction. Although this work is highly promising, it relies on computationally expensive sound source separation to obtain the percussive and harmonic parts of the signal, which increases training time. 

Previous work on self-supervised pitch estimation, SPICE \cite{spice2020}, introduced the concept of estimating relative pitch by fitting two shifted slices of a constant-Q transform (CQT) to the same convolutional autoencoder, and trained the encoder so that the difference in its outputs would be proportional to the introduced shift. This pretext task led to very good results while keeping pre-processing to a minimum: the computation of a CQT and the shift. 

Similarly to how a magnitude spectrogram or CQT represents time-frequency varying contents of a signal which are important for pitch analysis, the tempogram representation indicates for each time frame the local relevance of a tempo estimate for a given music recording \cite{muller2015chap6}. This makes it a compelling case to adapt an architecture such as SPICE to tempo estimation using a tempogram as input representation. On this basis, we explore the adaptation of SPICE to tempo estimation from a tempogram and study the different design choices as well as their effect on tempo estimation\footnote{At the time of submitting this paper a related and concurrent paper \cite{quinton2022equivariant} became available online on September 3rd, 2022.}.



\section{Self-supervised tempo estimation}
\label{sec:method}

\subsection{Tempo representation}
\label{ssec:tempo_rep}

Most tempogram representations are calculated in two stages \cite{muller2015chap6}. First, a novelty function is calculated, 
and second, the periodicity of the novelty function is analyzed to obtain a local tempo salience map. During this analysis, one of the main challenges relates to the multiple periodicities that exist in the hierarchical rhythmic structure of music, which are reflected in the tempogram as harmonics or sub-harmonics. Although this ``harmonic'' organization resembles harmonic-sound structures in a spectrogram representation, it can be more ambiguous, as the predominantly perceived tempo of a piece might have a low salience in the tempogram while its harmonics (or sub-harmonics) might have strong salience. This ``harmonic structure'' depends on the music and the method used to compute the tempogram (e.g autocorrelation vs. Fourier analysis). 

In order to understand the effect that these multiple periodicities have in our self-supervised framework, we compare the performance of the model using three tempogram formulations: an autocorrelation tempogram $\T{A}$~\cite{acf_notebook}, a Fourier-based tempogram $\T{F}$~\cite{dft_notebook},
and a \textit{hybrid} tempogram representation $\T{H}$~\cite{peeters2006template} that combines the previous two. To do so, we compute a novelty function using a single-band spectral flux and calculate $\T{A}$ and $\T{F}$ from it. 
Then, we convert $\T{A}$ from a time-lag representation to a time-tempo representation~\cite{acf_notebook}. 
This conversion results in nonlinear tempo values because of their correspondence to the linearly sampled lag values. To obtain an autocorrelation tempogram $\T{A}$ with the same tempo set as $\T{F}$, we interpolate the tempo values~\cite{muller2015chap6}. 
Finally, we compute $\T{H}=\T{A} \times \T{F}$.



Changes in frequency are perceived differently depending on the frequency range they happen, e.g. an increase of $40$~Hz from $40$~Hz to $80$~Hz is perceived as an octave, but an increase of $40$~Hz from $120$~Hz to $160$~Hz is perceived as a perfect fourth. Similarly, tempo changes are perceived relative to their speed \cite{grosche2010cyclic, kurth2006cyclic, gratton2016absolute}, so an increase of $30$ beats per minute (BPM, bpm) from $60$~bpm to $90$~bpm is perceived as a bigger change than $130$~bpm to $160$~bpm. With this in mind, we convert the tempo axis of the tempogram representations from a linear to a logarithmic scale, by combining the linear bins into logarithmic ones centered at 
\begin{equation}
    t_k = t_{0}2^\frac{k}{Q},
    \label{eq:log_index}
\end{equation} 
\noindent with $t_0=25$~bpm and $Q=40$ bins per octave. This bears a resemblance to how a CQT (which has a log-frequency axis) can be derived from a (linear-frequency) spectrogram.

\subsection{Self-supervised framework}
\label{ss:dnn}

\begin{figure}[htb]
  \centering
  \includegraphics[width=\linewidth]{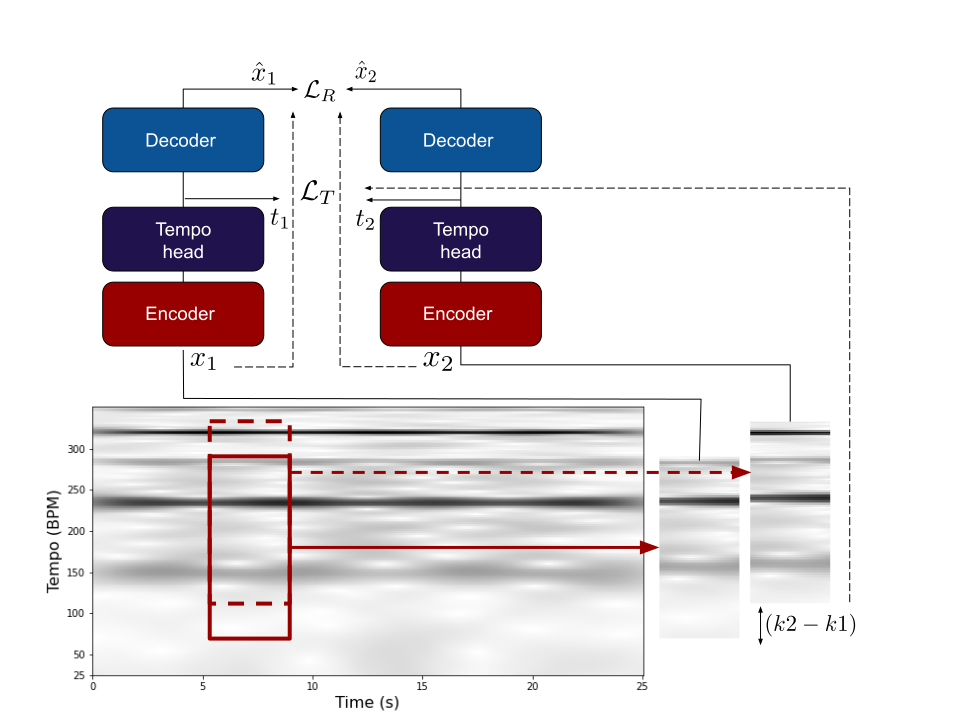}
  \caption{Overview of the training framework.}
  \label{fig:framework}
\end{figure}

Our self-supervised framework\footnote{Our implementation is open source and available at \url{https://github.com/giovana-morais/steme}} (Figure \ref{fig:framework}) is inspired by SPICE
\cite{spice2020}. 
Given a tempogram, our method extracts two random $128$-dimensional slices $x_{1}$ and $x_{2}$ which correspond to the same time instant but are shifted vertically by $k_1$ and $k_2$ bins respectively. These shifts translate to artificial changes in tempo in the different slices fed to the model, as the lines of the tempogram are vertically displaced. We sample the values $k_1$ and $k_2$ from the uniform distribution $\mathcal{U}(11, ..., 18)$, so the tempo range seen by the model ranges from 30~bpm to 310~bpm, i.e. for $k=11$ in Eq.~\ref{eq:log_index} the slice starts at $t_{11} = 25 \cdot 2^{\frac{11}{40}} = 30.2$~bpm, and ends at $t_{127+11} = 25 \cdot 2^{\frac{(127 + 11)}{40}} \approx 273.2$~bpm; similarly, for $k=18$ the input slice covers the interval $[34.1, 308.4]$~bpm.

As in \cite{spice2020}, we define the relative tempo error as 
\begin{equation}
    e_t = |(t_1-t_2) - \sigma (k_2-k_1)|,
    \label{eq:et}
\end{equation}

\noindent so the difference between estimated tempi in each branch has to be proportional to the artificially introduced difference $(k_2-k_1)$. The constant $\sigma$ ensures that the difference in the output model $t_1-t_2$ corresponds to the right number of bins within the training set, and is set to 
\begin{equation}
    \sigma=\frac{1}{Q\log _2\left(\frac{t_{\max}}{t_{\min}}\right)}
\end{equation} 
\noindent following~\cite{spice2020}, where $t_{\max}$ and $t_{\min}$ are the maximum and minimum tempi present in the training set, and $Q$ is the number of bins per tempo octave. We set $Q=40$ to ensure a reasonable tempo range and resolution, e.g. four tempo octaves 30-240~bpm correspond to $160$ bins. Unlike \cite{spice2020}, we observed that $k_2-k_1$ (instead of $k_1-k_2$) preserves the monotonic correspondence between BPM and model output values, as higher shifts correspond to tempogram slices with lower salience lines and thus lower tempi. 

The model is then trained to estimate the relative tempo within each slice by applying the Huber norm of the tempo error, resulting in the following loss function: 
\begin{equation}
    \mathcal{L}_T = \frac{1}{T}\sum_t h(e_t),
\label{eq:Lt}
\end{equation}
where $T$ is the batch size, $h$ is defined as
\begin{equation}
   h(x) = \begin{cases}
    \frac{x^2}{2},\; & \text{if}\ |x| \leq \delta, \\
    \frac{\delta^2}{2} + \delta(|x| - \delta),\; &\text{otherwise},
    \end{cases}
\end{equation}
\noindent and we set $\delta = 0.25$.

Besides the loss in Equation \ref{eq:Lt}, analogously to \cite{spice2020} we use a reconstruction loss as follows:

\begin{equation}
    \mathcal{L}_R = \frac{1}{T}\sum_T ||x_1-\hat{x}_1||^2 _2 + ||x_2-\hat{x}_2||^2 _2,
\label{eq:LR}
\end{equation}

\noindent where $x_1, x_2$ are the input slices to each model branch and $\hat{x}_1, \hat{x}_2$ are the reconstructed slices at the output of the decoder in Figure \ref{fig:framework}. The final combined loss is given by:

\begin{equation}
    \mathcal{L} = \omega_T \mathcal{L}_T + \omega_R \mathcal{L}_R,
\label{eq:Loss}
\end{equation}

\noindent with $\omega_T=10^4$ and $\omega_R=1$ (as in \cite{spice2020}).

We use a $6$-layered convolutional encoder that receives a $128$-dimensional vector corresponding to a slice of the tempogram and outputs one scalar representing tempo. We use filters of size $3$ and stride equal to $2$, and the number of channels is equal to $d \cdot [1, 2, 4, 8, 8, 8]$, where $d = 64$, and we use ReLU as the activation function for all layers. The output of the last convolutional layer is flattened and fed into a tempo estimation head that consists of two fully-connected layers with $48$ and $1$ units respectively. During training, the tempogram frames of the input audio are shuffled to ensure that different tempi are seen for a given batch. The scalar output of the encoder is then fed into a decoder which has the objective of reconstructing the tempogram slice from this tempo estimation. The decoder also has $6$ layers as $d \cdot [8, 8, 8, 4, 2, 1]$ and is composed of transposed convolutional layers, with the same filters, stride values, and activation function as the encoder. The weights between the two branches of the network are shared.

Unlike \cite{spice2020}, we do not have a confidence/voicing head and we do not make use of MaxPooling or BatchNormalization layers as informal testing revealed the inclusion of these steps was detrimental to learning.

\subsection{Model calibration}
The output of the tempo head is a real-valued scalar $t \in [0, 1]$, which indicates the estimated tempo within a given slice. To perform inference with this model, we need to map this output to a BPM range using a calibration step \cite{spice2020}. Considering the logarithmic nature of the tempogram, given a linear range of BPM values, the model output will have a logarithmic distribution of values, with the values for higher tempi being squeezed into closer activation values of $t$. On this basis, we generate synthetic metronome clicks following a logarithmic tempo range and map the model to BPM with a linear model. 


\section{EXPERIMENTS}
\label{sec:exps}

We investigate the impact of the tempogram input representation, the effect of the data distribution, and the interplay between the two in the context of self-supervised tempo estimation. To do so, we focus this study on the analysis of synthetic data and perform controlled experiments by training nine models, one for each combination of data distribution and tempogram. We also train one model for each tempogram representation using real data as a comparison point. 


\subsection{Synthetic data generation}
\label{ssec:data_dist}

\begin{figure}[htb]
   \centering
   \includegraphics[width=\linewidth]{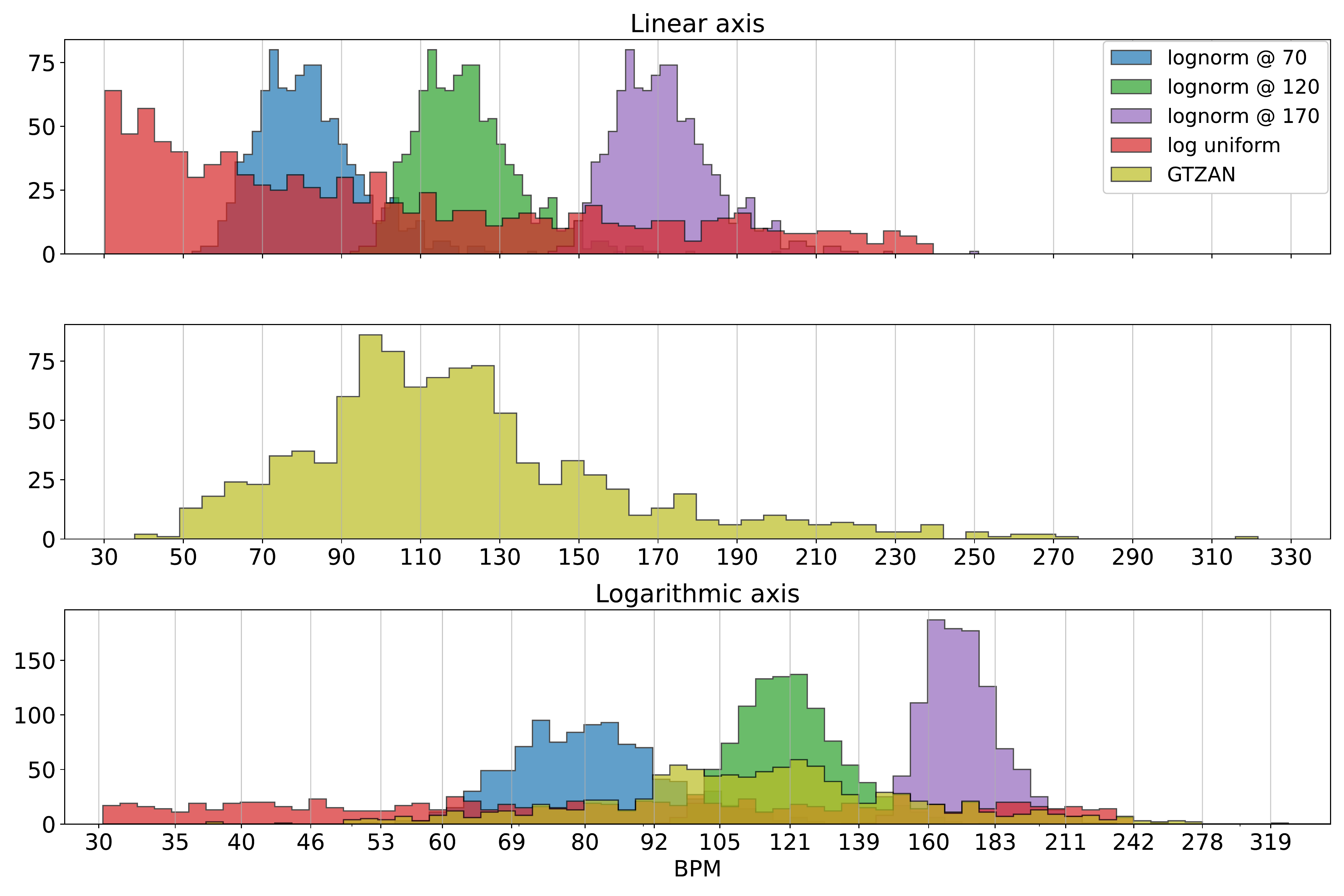}
   \caption{Synthetic data and GTZAN distributions. The first row shows synthetic data distributions over a linear axis. The second row shows the GTZAN data distribution over a linear axis. The third plot shows all data distributions over a logarithmic axis. It is possible to note that, for the logarithmic axis (which corresponds to the lines of the tempograms), distributions centered on higher tempi have narrower support than the distributions centered on lower tempi.
   }
   \label{fig:tempo_dist}
\end{figure}

To understand how the distribution of the data affects the self-supervised model performance, 
we synthesize metronome excerpts ($N=1000$) following four different distributions. For three, we model tempo as a \textit{log-normal} random variable $X$ such that $\log_2(X) \sim \mathcal{N}(\mu, \sigma^2) $, with $\mu \in \{70, 120, 170\}$~BPM and $\sigma = 0.25$. These three distributions are modeled after real-world datasets (e.g. GTZAN~\cite{tzanetakis2002gtzan, marchand2015gtzanrhythm}) shifted to different tempo ranges. We also include a \textit{log-uniform} distribution where $\log _2 (X) \sim \mathcal{U}([30, 240])$, representing a well-balanced ideal tempo distribution, which is difficult to find in practice but serves as an upper limit reference for our experiments. See resulting distributions in Figure \ref{fig:tempo_dist}.


\begin{figure*}[ht!]
   \centering
   \includegraphics[width=\linewidth]{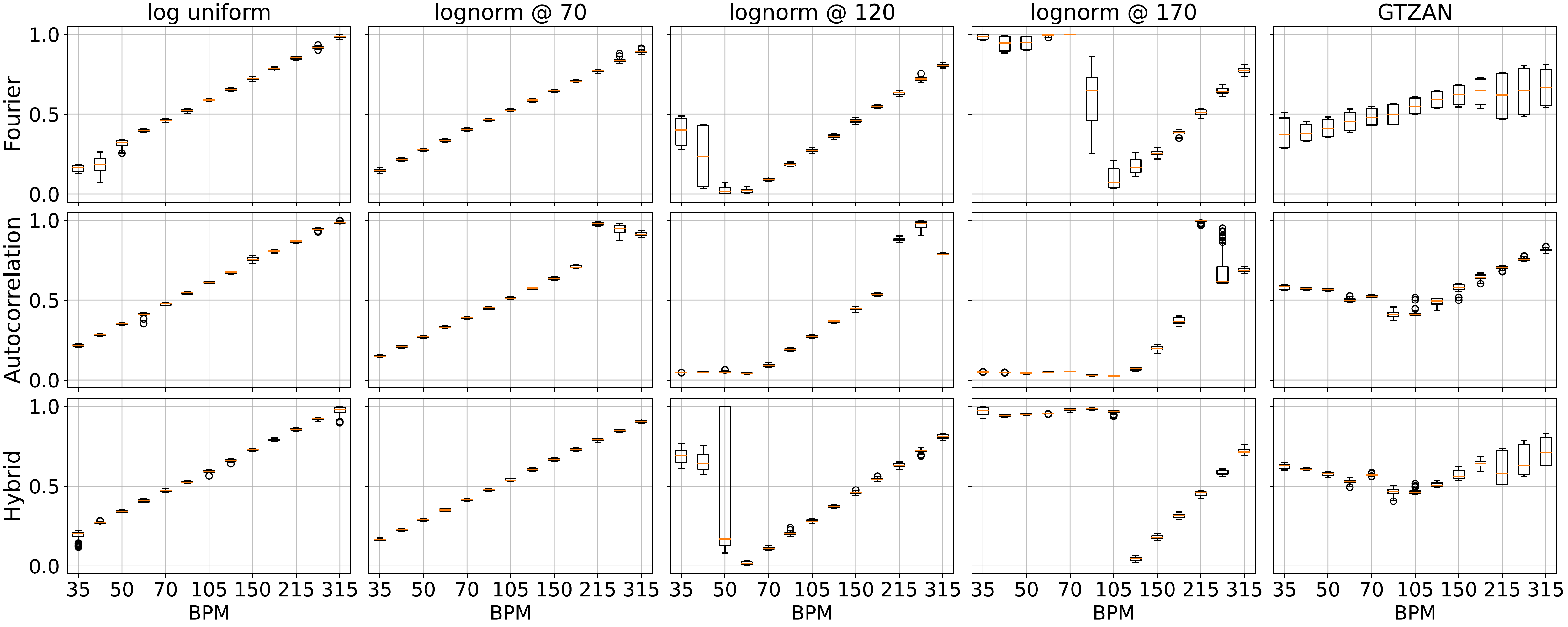}
   \caption{Model calibration results for different data distributions and tempogram representations. All x and y axis are shared between plots.}
   \label{fig:main_result}
\end{figure*}

\subsection{Implementation and evaluation}

The model is implemented in Keras \cite{chollet2015keras} and trained using Adam \cite{kinga2014adam} with default hyperparameters and a learning rate equal to $10^{-4}$. The batch size is $64$ and all the experiments were trained for $15$ epochs. The spectral flux used in all tempograms was calculated using librosa~\cite{librosa}, with a window size of $2048$ and hop length of $512$ samples for audio sampled at $22$~kHz. We apply a logarithmic compression with $\gamma = 100$ and normalize the resulting signal. For the tempogram, we use a $10$-second window and hop length of $1$ spectral flux sample ($\sim$ $23.2$~ms). The tempogram implementations were based on libfmp~\cite{libfmp}.

All synthetic data consist of $60$-second metronome tracks that follow different tempi distributions. The log-normal distributions were generated using SciPy 
\cite{scipy} and the log-uniform was defined by 
$t_{\min}\cdot e^{\log{\left(i\cdot\ t_{\max}/t_{\min}\right)}}$, where $i \in [0,1)$.





\subsection{Results and discussion}
\label{ssec:results}

Since calibration results correlate with model performance, i.e. a model with more linear and less disperse calibration curves leads to more consistent estimations \cite{spice2020}, we look at calibration results for each model, depicted in Figure \ref{fig:main_result}.


The Fourier tempogram shows more consistent calibration results across data distributions and leads to smoother curves for both synthetic and real data. This result aligns with the intuition that the Fourier tempogram is the most similar to the CQT representation in \cite{spice2020} as it tends to have upper harmonics but not sub-harmonics. On the contrary, the autocorrelation tempogram results present the most variation, meaning that for similar tempo values, the model struggles to assign the same tempo output. Surprisingly, the hybrid tempogram does worse than the Fourier tempogram, 
which suggests that the multiplication of the autocorrelation and Fourier tempograms removes harmonic information that the model relies on to perform tempo estimation. 

The experiments with the different data distributions show that the model does not benefit from the wider range of a log-uniform distribution, and instead is able to extrapolate from a log-normal distribution centered in the lower tempi end ($70$~bpm). When we look at this result closely, an explanation may be found in Figure \ref{fig:tempo_dist}. Given a tempo distribution expressed in linear BPM values (Figure \ref{fig:tempo_dist} top), the model will be fed a slightly different tempo distribution (Figure \ref{fig:tempo_dist} bottom) because of the logarithmic effect of the input tempogram. This logarithmic representation compresses the tempo range increasingly in higher tempi, meaning that the model actually has access to a smaller range of log BPM values. In more extreme cases, such as the log-normal distribution centered at $170$~bpm, the model learns to use the whole range of its output for the rather small range of tempo values it sees, and ``saturates'' for values out of range, i.e. it assigns the same value to all unseen tempi. This suggests that it would be more beneficial to annotate or augment data in this lower range of tempi to ensure a more stable model.

Finally, when we look at the results for the model trained with real data, we see that the same observations hold. The Fourier tempogram leads to a more stable model, which shows a larger variance than models trained with synthetic data due to the inherent variability and noisiness of real data which complicates training. The GTZAN distribution has similarities with the lower-bpm log-normal distributions, which shows in the smooth trend of Figure \ref{fig:main_result} with the Fourier tempogram representation. For the GTZAN 
data, the hybrid tempogram also does worse than the other two variations, which suggests that we might introduce artifacts and noise by combining the two tempograms without further processing of the signal, as done in \cite{peeters2006template}, even though this is enough for synthetic data given its simplicity and steadiness.



\section{CONCLUSIONS}

We 
investigated the interplay between the tempogram input representation and the data distribution for self-supervised tempo estimation using synthetic data, 
demonstrating the challenges and considerations for adapting SPICE from pitch to tempo, as opposed to trying to directly advance the state-of-the-art in tempo at this stage. We found that the Fourier tempogram led to more consistent and precise models and that when observed BPM values were concentrated as real data distributions are, distributions centered around lower tempo values led to more successful models. In future work, we will explore these findings in the context of real data, and use these insights for guiding data augmentation.

\section{ACKNOWLEDGMENTS}
\label{sec:acks}

Thanks to the IEEE Signal Processing Society and the ME-UYR Program. The third author is funded by CNPq Grant 310141/2022-2. We also wish to thank F\'{e}lix Davies for the fruitful discussions. Finally, we would like to thank the anonymous reviewers for their comments and feedback on this work. 


\bibliographystyle{IEEEbib}
\bibliography{strings,refs}

\end{document}